\documentclass[sigconf, nonacm]{acmart}

\AtBeginDocument{%
  }

\usepackage{graphicx}
\usepackage{booktabs}
\usepackage{array}
\usepackage{CJKutf8}
\usepackage{tgtermes}

\begin{document}

\title{Animating Language Practice: Engagement with Stylized Conversational Agents in Japanese Learning}

\author{Zackary Rackauckas}
\email{zcr2105@columbia.edu}
\affiliation{%
  \institution{Columbia University, RoleGaku}
  \city{New York}
  \state{New York}
  \country{USA}
}

\author{Julia Hirschberg}
\email{julia@cs.columbia.edu}
\affiliation{%
  \institution{Columbia University}
  \city{New York}
  \state{New York}
  \country{USA}
}


\begin{abstract}
We explore \textit{Jouzu}, a Japanese language learning application that integrates large language models with anime-inspired conversational agents. Designed to address challenges learners face in practicing natural and expressive dialogue, \textit{Jouzu} combines stylized character personas with expressive text-to-speech to create engaging conversational scenarios. We conducted a two-week in-the-wild deployment with 52 Japanese learners to examine how such stylized agents influence engagement and learner experience. Our findings show that participants interacted frequently and creatively, with advanced learners demonstrating greater use of expressive forms. Participants reported that the anime-inspired style made practice more enjoyable and encouraged experimenting with different registers. We discuss how stylization shapes willingness to engage, the role of affect in sustaining practice, and design opportunities for culturally grounded conversational AI in computer-assisted language learning (CALL). By framing our findings as an exploration of design and engagement, we highlight opportunities for generalization beyond Japanese contexts and contribute to international HCI scholarship.
\end{abstract}



\begin{teaserfigure}
  \centering
  \includegraphics[width=\textwidth]{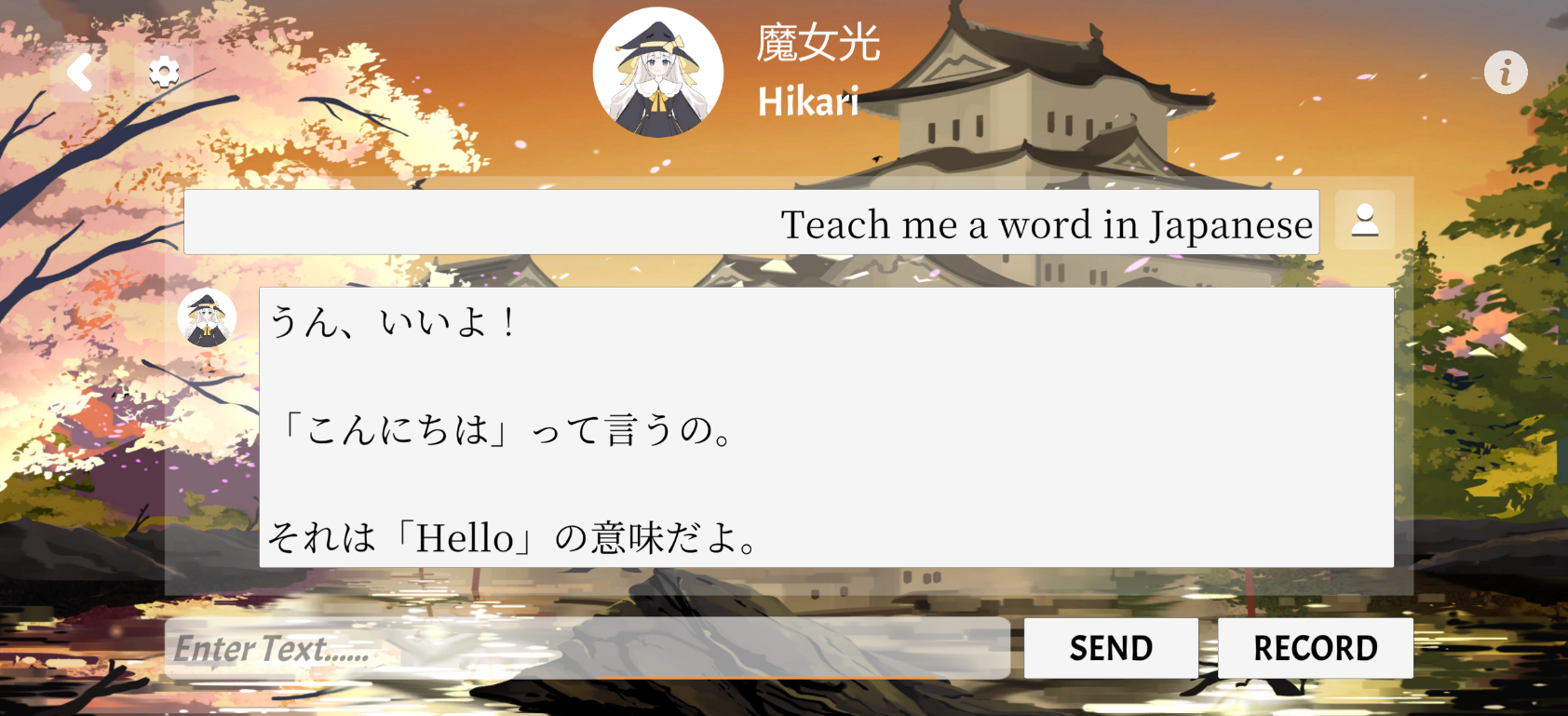}
  \caption{A screenshot from our Japanese language learning mobile app, \textit{Jouzu}, featuring “Hikari,” an anime-inspired conversational agent. Users interact via text ("send") or voice ("record"), receiving stylized, in-character responses designed to enhance engagement and cultural immersion.}
  \Description{Screenshot of a conversation between the user and an anime-style agent named Hikari. The user asks for a Japanese word, and the agent replies in Japanese with a stylized, character-specific response.}
\end{teaserfigure}

\maketitle

\section{Introduction}

As conversational agents become integrated into everyday tools, the stylization of an agent's voice, character, and aesthetics has become essential to the user experience. In human–computer interaction (HCI), conversational agents are not just tools, but social actors. They express personality, mediate emotion, and structure interaction in ways that influence user trust, engagement, and learning. Stylization is therefore a central design dimension with implications not only for education, but also for how people engage with technology in culturally specific and affectively rich ways across diverse domains.

At the same time, learning Japanese presents challenges reflected in both recent and foundational scientific literature. For example, it is ranked among the most difficult languages for native English speakers to learn, requiring approximately 2,200 classroom hours to reach proficiency \cite{usdos_fsi_foreign_language_training}. Beyond its multiple writing systems, Japanese diverges significantly from English in grammar \cite{boboc2023} and phonology \cite{ohata2004}. Traditional tools such as textbooks are often rigid, impersonal, and lack spoken interaction, contributing to learner disengagement and high dropout rates. Conversational practice remains one of the most effective techniques for language acquisition, yet it is often limited by social anxiety and access to native speakers \cite{li2022}. These constraints position language learning as a high-barrier context in which the role of agent stylization may be particularly impactful.

We focus on providing a multimodal interface where Japanese learners engage with fictional agents designed to emulate recognizable character archetypes from {\it anime}, a form of Japanese animation that has become ingrained in global pop culture. To provide fun, engaging, and personalized learning to students who may otherwise quit due to the high barriers to entry of Japanese, we conducted an in-the-wild study using the \textit{Jouzu} mobile app, which leverages large language models (LLMs) and an expressive multilingual text-to-speech (TTS) system using Style-BERT-VITS2 \cite{rackauckas2025benchmarkingexpressivejapanesecharacter}. Unlike prior computer-assisted language learning (CALL) systems that rely on neutral chatbots or scripted tutors, produces emotionally rich, character-specific Japanese text and speech responses in conversational agents. These agents employ language with non-standard pronouns, stylistic particles, and expressive phrasing to simulate social presence and emotional realism \cite{kinsui2015}. They are paired with anime-inspired visuals and animations to create an environment that combines cultural familiarity with affective depth. Through these features, the agents act as narrative participants who provide a stress-free and engaging conversational environment rather than a traditional, robotic chatbot tutor.

As a whole, \textit{Jouzu} combines such conversational characters with prescripted QA dialogues, referred to as "lessons," an in-text sentence-level inspector that provides word-by-word definitions and readings and flashcards to practice vocabulary from both lessons and conversation created with the in-text inspector. These multimodal conversational environments are designed to replicate human interaction without judgment, thereby encouraging playful exploration in chats alongside structured practice in lessons and flashcards. The visuals of the app, including flashcards, are grounded in anime stylization, tying the system’s linguistic playfulness to a coherent aesthetic world.

In a mixed-methods study with 52 participants, we investigated how users of varying proficiency and linguistic backgrounds interacted with the application. We collected quantitative ratings of usability, engagement, and agent perception, alongside open-ended feedback and behavioral data from chatbot logs. Our analysis describes patterns in how agent stylization including voice, visual design, and character behavior related to reported user experience, emotional responses, and interaction strategies. In some cases, we also tested for significant differences between user subgroups (e.g., native language and proficiency level), allowing us to examine potential impacts of linguistic background on app usage. This study therefore not only contributes to the design of language learning systems, but also extends ongoing discourse in HAI and HCI on the role of personality, culture, and embodiment in agent design.

We have three contributions:
\begin{enumerate}
\item A novel conversational learning system that combines large language models, a style-transfer TTS engine (Style-BERT-VITS2), and anime-inspired personas to support second-language (L2) practice.
\item An in-the-wild deployment study with 52 Japanese language learners, offering empirical insights into how stylized conversational agents influence learner engagement, motivation, and willingness to experiment with language in everyday contexts.
\item Design implications for HCI and CALL, highlighting how cultural stylization and affective expression can support engagement and creative language use, and discussing the opportunities and challenges of deploying such systems across different learner backgrounds and proficiency levels.
\end{enumerate}

What distinguishes this work is its exploration of stylization as more than surface aesthetics, examining how anime-inspired characters, expressive speech synthesis, and culturally situated design elements interact with learners across diverse backgrounds. More broadly, this study contributes to HCI by showing how culturally resonant, affective stylization can be leveraged to design agents that are not only functional but also socially and emotionally meaningful. This positions stylization as a key lever for designing next-generation conversational agents in HCI, where personality and culture are as central as functionality.

\section{Related Work}

Early HCI research established principles that remain directly relevant to conversational agents for language learning. Card, Moran, and Newell’s keystroke level model provided a rigorous framework for predicting expert, error free task performance and showed how even small interaction costs shape usability \cite{Card1980}. While language learners are novices rather than experts, this insight underscores the need to minimize friction in conversational systems. Furnas et al. highlighted the vocabulary problem and demonstrated that users often employ diverse and mismatched terms, a challenge that remains central to natural language interfaces and learner input variability \cite{Furnas1987}. Grudin emphasized that computing is inherently social and collaborative and framed systems as situated within organizational and cultural contexts. This insight resonates with the social nature of language learning \cite{Grudin1994}. Shneiderman articulated core usability principles of consistency, informative feedback, and error prevention and handling in his book {\it Designing the User Interface}. These principles continue to underpin effective CALL design \cite{Shneiderman1986}. Our work extends these foundations by exploring how stylized voiced characters draw on cultural and affective dimensions to support engagement and learning in multimodal CALL environments.

Language learning systems increasingly incorporate conversational agents to simulate spoken practice in a low-risk environment. Traditional computer-assisted language learning (CALL) tools have relied on scripted responses and task-based exercises, but newer systems use LLMs to support open-ended dialogue, generate feedback, and scaffold instruction \cite{cao2023, dornburg2024}. Question-answer chatbots have been deployed in multiple languages to provide real-time practice, although most remain text-based \cite{li2022}. More recent work has incorporated empathetic language to reduce anxiety and increase user retention \cite{siyan2024a, siyan2024b}. Aiba et al. \cite{aiba2024} introduced a system in which English learners could practice conference-style questions with GPT-generated speech, suggesting that combining TTS with LLMs encourages natural spoken interaction.
Voiced characters extend this trend by merging LLM dialogue with expressive speech synthesis and cultural stylization. While most prior systems employ neutral or utilitarian speech, \textit{Jouzu} introduces character-specific language rooted in anime tropes --offering linguistic diversity and affective realism. These characters use “character language” featuring stylized pronouns, emphatic particles, and emotionally expressive phrasing, contributing to both social presence and learner motivation.

Kinsui and Yamakido \cite{kinsui2015} describe “character language” as a stylized form of Japanese used in fiction to signal personality, gender, or origin. These features include non-standard pronouns, archaic verb forms, and distinctive particles. Although rare in textbooks, they are recognizable and emotionally charged for many learners. Li et al. \cite{li2023haruhi} explored this concept with ChatHaruhi, a chatbot modeled on anime characters, but did not include voice interaction.
In contrast, \textit{Jouzu} introduced voiced agents with character-specific linguistic stylization rooted in anime tropes. These agents use “character language” \cite{kinsui2015} to emulate fictional personas with affective realism. This represents a shift from purely pedagogical interactions toward immersive, socially engaging dialogue that blends linguistic and cultural learning.

Multimodal interfaces are central to current human-computer interaction (HCI) discussions about language learning. OpenAI's GPT-4o \cite{openai2024gpt4technicalreport}, for example, enables faster response generation with integrated audio \cite{ying2024}, supporting smooth, responsive dialogue. At the same time, systems must address concerns about equity, safety, and engagement. Chan and Tsi \cite{chan2023} note that computer-assisted tutors must maintain educational integrity while sustaining user interest. Xue et al. \cite{xue2023} highlight the importance of filtering harmful outputs in generative systems.

Previous HCI research has leveraged user studies to investigate and proposed guidelines for interactions with learner dialogue systems. For example, through a user study of 107 users, Koyoturk et al. assert the need for effective prompting in learner chatbot interactions, highlighting the need to provide user guidelines to learn how to prompt the agents \cite{koyuturk2025understandinglearnerllmchatbotinteractions}. Cai et al. conducted a user study with a group of three undergraduate students to advance dialogue system understanding in small group learning \cite{caietal}. Annamalai et al. conducted a user study with 360 undergraduate students to investigate chatbot language learning in Malaysian higher education institutions \cite{Annamalai2023Chatbots}. Finally, Li et al. conducted a user study with 24 students to assess a curriculum-driven chatbot to match students' classroom curricula \cite{li-etal-2024-curriculum}.
Similarly, we employ a user study with 52 users to analyze user interactions and learning outcomes after their use of \textit{Jouzu}, centered around the conversational character agents.

\section{System Architecture}

\textit{Jouzu} integrates voiced character agents, scripted dialogues, low-stakes stylized quizzes, and an interactive sentence-level inspector for building vocabulary flashcards. While flashcards and scripted lessons are familiar features in computer-assisted language learning, \textit{Jouzu}'s design emphasizes stylization and immersion, elevating these methods into a playful and affectively engaging environment.

\subsection{Original Characters}
\textit{Jouzu} features three original characters with distinct personalities, backstories, voices, and speaking styles. Each character was designed to represent recognizable anime tropes. In this study, the three characters deployed were a young girl attending a magic school who aspires to master light magic (Hikari), a Japanese kitsune deity capable of taking human form (Kitsune), and a samurai from Edo-period Japan. Each character employs a distinctive character language style (i.e., a casual young female, a wise centuries-old female deity, and a masculine-coded samurai), providing learners with exposure to diverse forms of Japanese vocabulary, grammar, and expressive speech.

These characters coexist within a stylized fictional world that blends elements of magic, prayer, and samurai culture. Professional character and background art reinforce immersion, grounding the linguistic stylization in a coherent, affectively rich environment. The novelty lies in how these fictionalized archetypes extend traditional CALL agents by embedding language practice within culturally familiar yet emotionally evocative anime-inspired worlds.

\subsection{Dialogue System Architecture}
The conversational character system in \textit{Jouzu} was designed to move beyond standard chatbot architectures. Natural conversations between humans and GPT-4 were first leveraged to create QA scripts. Each script was divided into ten sections, with each section containing three to five turns between the user and the LLM. These QA scripts provided the foundation for lessons and informed the training of multimodal text and text-to-speech models.
This design highlights a methodological contribution: by grounding conversational scaffolds in pre-authored human–LLM dialogues, \textit{Jouzu} ensures pedagogical coherence while retaining the flexibility of real-time generation.

\subsubsection{Text Generation}
During real-time, user-driven interaction, an LLM (e.g., GPT-5) is prompted with contextual cues and instructions to emulate each character’s distinct speaking style. This ensures that user exchanges remain consistent with the intended persona and stylization of the character. 
The contribution here is in demonstrating how real-time prompting for stylized persona-specific language can extend beyond neutral tutoring into affectively rich, culturally anchored interaction.

\subsubsection{Speech Synthesis}
\textit{Jouzu} employs Style-BERT-VITS2 for speech synthesis. Professional Japanese voice actors recorded the QA scripts in each character’s voice, capturing distinct character language and stylistic nuance. These recordings were used to fine-tune models on top of the pretrained Style-BERT-VITS2 JP Extra model. Relevant work shows that Style-BERT-VITS2 JP Extra produces synthesized speech that, on average, shows no significant difference from native Japanese ground truth \cite{rackauckas2025benchmarkingexpressivejapanesecharacter}. This integration of professional voice acting with state-of-the-art synthesis contributes to HCI by demonstrating how affectively expressive, near-indistinguishable synthetic speech can be deployed in real-world language learning contexts.
\begin{figure*}[h]
  \centering
  \includegraphics[width=\textwidth]{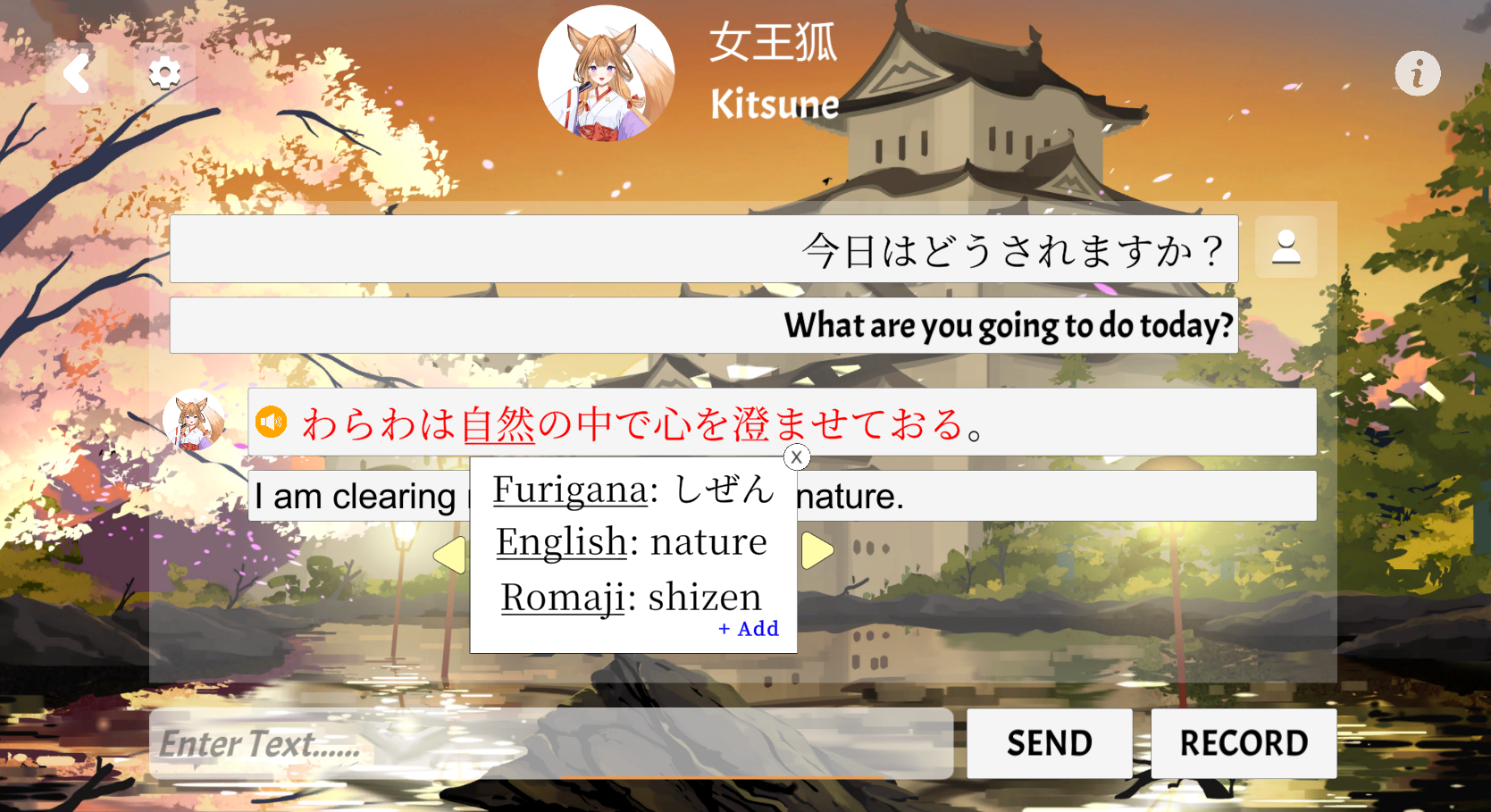}
  \caption{A screenshot from \textit{Jouzu} featuring the in-text word inspector. The image shows a sent user message with the translation, "What are you going to do today?" and a response from the character Kitsune. The user has tapped on the sentence, and the flashcard for the Japanese word "shizen," translated to "nature" in English, appears with the furigana, English definition, and romaji spelling of the word.}
  \label{fig:text-inspector}
  \Description{A screenshot from \textit{Jouzu} featuring the in-text word inspector. The image shows a sent user message with the translation, "What are you going to do today?" and a response from the character Kitsune. The user has tapped on the sentence, and the flashcard for the Japanese word "shizen," translated to "nature" in English, appears with the furigana, English definition, and romaji spelling of the word}
\end{figure*}

\subsection{In-text Inspector}
Japanese learners, particularly those from non-Chinese language backgrounds, often struggle with reading due to the stark differences from other languages in grammar \cite{boboc2023} and phonology \cite{ohata2004}. To address this, \textit{Jouzu} includes an in-text inspector: during any chat or lesson, users can click on a sentence to trigger flashcards that define each word step-by-step, including readings and pronunciations (see Figure \ref{fig:text-inspector}).
The novelty of this feature lies in its seamless integration of word-level scaffolding into conversational dialogue, offering immediate, interactive support without breaking immersion.

\section{Method}

A user study was conducted with 54 participants initially recruited; of these, 52 fully completed all parts of the study, including surveys and app usage. The analysis below focuses on these 52 complete cases to ensure consistency and validity. We examine app usability, learning outcomes, engagement, impact of stylized characters on the learning experience, and  comparison of \textit{Jouzu} to other language learning apps like Duolingo. In addition, we gathered qualitative feedback on user experiences and analyzed interaction patterns between users and conversational agents. 

Participants were recruited from a university computer science class. They were then instructed to freely use the app for at least one hour within a period of two weeks. 
%jh: one hour per day or ??
In addition, we asked them to use the conversational agents present in the app. Users completed a survey consisting of multiple choice questions and open responses before (see Appendix \ref{appendix:survey_questions}) and after the app (see Appendix \ref{appendix:evaluation_questions}). The initial survey collected demographic and background information from users and presented an optional 10-question multiple choice Japanese quiz with an optional short paragraph response in Japanese (see Appendix \ref{appendix:quiz_questions}). The post-use survey asked several questions about the user experience, followed by a new 10-question Japanese quiz of the same difficulty level and an open response in Japanese (see Appendix \ref{appendix:quiz_questions}). Participants were rewarded with 5 assignment points in their university class. Of the 54 recruited participants, 52 fully completed the task.

All participants were students completing this work as part of a classroom assignment. Participation in assessments and activities was voluntary: students who chose not to participate did not incur any penalty, while those who did participate received extra credit. Emails were collected solely for the purpose of linking responses within the study; no identifying details are disclosed in publications, and all data were anonymized prior to analysis. Participation complied with our institution’s policies on classroom-based research activities. In conducting and reporting this work, we adhered to the ACM Policy on Research Involving Human Participants and Subjects, ensuring that participants’ autonomy, privacy, and well-being were respected.

\subsection{Descriptive Statistics of User Experience}

For quantitative analysis, Likert scale questions were grouped into important aspects for evaluation of the app. The following scale was used for groups one through 5:
\begin{enumerate}
\item 1: Not at all
\item 2: Slightly
\item 3: Somewhat
\item 4: Moderately
\item 5: Extremely
\end{enumerate}

For group six:

\begin{enumerate}
\item 1: Much worse
\item 2: Slightly worse
\item 3: The same
\item 4: Slightly better
\item 5: Much better
\end{enumerate}

\begin{enumerate}
    \item \textbf{Usability}
    \begin{itemize}
        \item How easy was the app to use?
    \end{itemize}

    \item \textbf{Amount Learned}
    \begin{itemize}
        \item How much did the app help you learn or practice Japanese?
    \end{itemize}

    \item \textbf{Fun and Engagement}
    \begin{itemize}
        \item How engaging was the app overall?
        \item How fun was it to use?
        \item How motivated did you feel to keep using it?
        \item How much did the animations enhance your engagement?
    \end{itemize}

    \item \textbf{Characters}
    \begin{itemize}
        \item How much did the characters enhance your learning experience?
        \item Which character did you enjoy chatting with the most?
        \item Did the characters make the app more enjoyable overall?
    \end{itemize}

    \item \textbf{Visuals, Art, and Aesthetic}
    \begin{itemize}
        \item How much did the anime-style visuals make the app more appealing?
    \end{itemize}

    \item \textbf{Comparison to Other Apps}
    \begin{itemize}
        \item Compared to other language learning apps, how would you rate the app in terms of fun?
        \item Compared to other language learning apps, how would you rate the app in terms of helpfulness for learning?
        \item Compared to other language learning apps, how would you rate the app in terms of ease of use?
        \item Compared to other language learning apps, how would you rate the app in terms of your motivation to keep learning?
        \item Compared to other language learning apps, how would you rate the app in terms of visual and auditory style?
    \end{itemize}
\end{enumerate}

\subsection{Hypotheses and Analysis Design}

We tested two hypotheses for significance:
\begin{enumerate}
    \item Native Chinese (Mandarin) users will learn more than users of other native backgrounds since Japanese and Chinese share Kanji characters.
    \item Users with a higher level of Japanese will learn more and find the app more fun and engaging compared to beginners of Japanese due to \textit{Jouzu's} immersive approach.
\end{enumerate}

To examine potential cross-linguistic differences in user experience, participants were grouped by native language: Chinese and non-Chinese speakers.
To assess the significance of these cross-linguistic differences, we conducted independent-sample t-tests comparing mean group ratings between Chinese and non-Chinese participants across all six experience categories.
We also grouped users by self-reported Japanese proficiency (Complete Beginner, Beginner, Intermediate, Native) to analyze mean ratings by language experience level. A heatmap was generated to visualize rating trends across these categories.

\subsection{Qualitative Data Collection}

Participants responded to four open-ended prompts:
\begin{itemize}
    \item What was the most helpful feature for learning Japanese?
    \item What was the least helpful feature?
    \item What was the most fun and engaging feature?
    \item What did you find frustrating or confusing?
\end{itemize}
Responses were mapped to one of six predefined experience categories: Usability, Amount Learned, Fun and Engagement, Characters, Visuals, Art and Aesthetic, and Comparison to Other Apps.

\subsection{Realized Learning Outcomes}

\subsubsection{Multiple Choice}
Users were asked ten multiple choice questions before and after use of the app. All questions were optional. The questions are included in Appendix \ref{appendix:quiz_questions}. Pre- and post-use completion rates and accuracy scores were compared to assess realized learning outcomes in terms of grammar and vocabulary.
Ten questions were asked. The first five questions included vocabulary sampled from the characters' individual language styles. The next five were questions sampled from each Japanese Language Proficiency Test Level (N1 through N5). 
The pre- and post-tests were different, but were created to be the same difficulty for respondents.

\subsection{Chatbot Interaction Logging}

User interaction data with the character-based chatbot was analyzed qualitatively.  Patterns were identified in beginner vs.~advanced usage, including:
\begin{itemize}
    \item English queries for vocabulary (e.g., “What is ‘sword’ in Japanese?”)
    \item Passive listening via English prompts with Japanese replies
    \item Full-sentence Japanese dialogue from intermediate and native speakers
    \item Use of speech input for pronunciation and listening practice
\end{itemize}
These patterns were coded from logged conversations or retrospective user descriptions.

\subsection{Usage Statistics}
To better understand patterns of engagement, we collected usage data via server-side timestamps. From this data, we derived the following quantitative measures:
\begin{enumerate}
    \item Average number of chat messages per user: total messages sent per participant, normalized by session count.
    \item Average number of sessions per user: discrete usage periods separated by at least 30 minutes of inactivity.
    \item Average session duration: computed from the first to last interaction within a session window.
    \item Most-used characters: relative frequency of interactions with each stylized agent (Hikari, Kitsune, Samurai).
\end{enumerate}
These measures provided a baseline of overall interaction with the application and allowed us to contextualize survey responses with actual engagement behavior.

% TODO: ADD CITATIONS FOR RUBRIC CREATION
\subsection{Writing Sample Evaluation (Rubric-Based)}

Participants were invited to write a short paragraph in Japanese both before and after using the app. This task was optional and open-ended, allowing users at any level to attempt expressive written output.
To evaluate changes in Japanese proficiency, we applied a rubric-based assessment focusing on vocabulary diversity, grammatical complexity, and stylistic features such as character language. 

These dimensions follow well-established frameworks in second-language research. Vocabulary diversity has long been recognized as a marker of language development, often measured through indices of lexical richness \cite{malvern2004lexical}. Grammatical complexity is similarly central, with studies showing that features like clause chaining and subordination track learner proficiency \cite{lu2010automatic}. Finally, we included stylistic and pragmatic appropriateness, drawing on the CEFR’s emphasis on register, politeness, and sociolinguistic control in real communicative settings \cite{council2001cefr}. Taken together, these categories reflect the broader complexity–accuracy–fluency (CAF) framework in applied linguistics \cite{housen2012dimensions}, ensuring that our rubric aligns with widely accepted standards for assessing learner language.

A near-native Japanese teacher independently scored all pre- and post-use paragraphs on a one, three, or five scale per dimension (higher is better). If no paragraph was written, zero was assigned for each category.
\ref{appendix:rubric}. We deliberately used one/three/five anchors to reduce ambiguity and emphasize categorical distinctions. This contrasts with the 5-point Likert items but improves coding reliability.

\section{Analysis}

\subsection{User Background}
In the pre-use survey, background information including native language, other languages spoken, age, and other language learning apps used was collected from users. Below, we present the background information of the users who completed the study and provided background information:  50 participants.
English was the most frequently reported native language, accounting for 19 responses, with bilingual variants such as Korean and English present. Other native languages included Chinese (17), Korean (3), Hindi (5), Tamil (1),  Turkish (1), Japanese (1), and Malayalam (1), reflecting a linguistically diverse sample. The age distribution showed a concentration of younger participants, with an average age of 24.32 years (SD = 3.78). The age range extended from 20 to 43 years, with 50\% of respondents aged 23 or younger and 75\% under the age of 25, indicating that most participants were in their early twenties. 39 participants had previously used Duolingo. Other language learning apps used included Rosetta Stone (4), Busuu (3), Babbel (3), and Memrise (1).

\subsection{Qualitative Analysis}

These descriptives establish a baseline for our core question: how stylization relates to user experience in a high barrier learning context. Summarizing group means shows where the system does work and where friction remains, and it motivates the subgroup tests by language background and proficiency as well as our triangulation with qualitative accounts. The dimensions align with our constructs: usability as a prerequisite for continued use, visuals and characters as stylization, fun and engagement as motivation, and perceived learning as an outcome-adjacent perception that can diverge from engagement.

Participant responses were aggregated across six experience dimensions. Usability received the highest average rating (\textit{M} = 3.64), followed closely by Visuals, Art, and Aesthetic (\textit{M} = 3.52) and Characters (\textit{M} = 3.39), indicating that users responded positively to the interface design and stylized agents. Fun and Engagement also scored well overall (\textit{M} = 3.31, \textit{SD} = 0.36), though variation in motivation and enjoyment was observed. Comparison to other apps showed moderate ratings (\textit{M} = 3.10, \textit{SD} = 0.46), reflecting mixed perceptions of how \textit{Jouzu} compares to traditional language learning tools. Perceived learning outcomes were lowest (\textit{M} = 2.48), suggesting that while the app was enjoyable and visually engaging, users may not have viewed it as highly effective for structured language acquisition. 

\begin{figure}[h]
  \centering
  \includegraphics[width=0.95\linewidth]{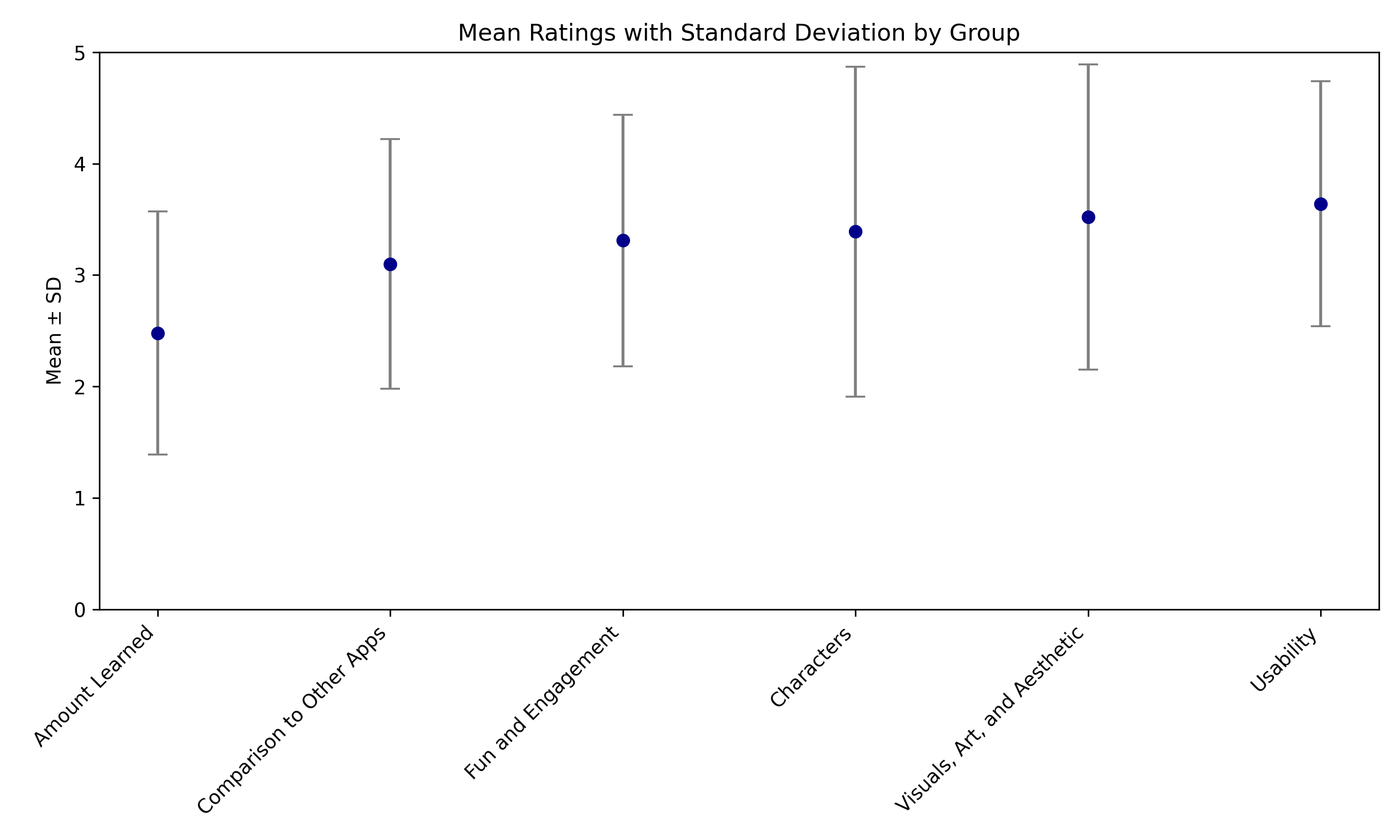}
  \caption{Mean participant ratings by experience category with standard deviation error bars (Likert scale: 1–5).}
  \Description{A grouped bar chart showing mean participant ratings by experience category on a Likert scale from 1 to 5. Each bar represents an experience category, and error bars indicate the standard deviation of ratings within that group.}
  \label{fig:grouped_bar_means}
\end{figure}

Figure~\ref{fig:grouped_bar_means} shows the average rating per group with standard deviation error bars, highlighting the overall trends and relative spread between each category. Table~\ref{tab:group_summary} presents the full descriptive statistics by group.

\begin{table}[h]
\centering
\caption{Group-level descriptive statistics of user ratings (Likert scale: 1–5)}
\Description{A table summarizing user ratings across six groups. Usability and Visuals received the highest mean ratings, around 3.5 to 3.6, while Amount Learned had the lowest mean rating at 2.48. Standard deviations are reported where applicable, with variability ranging from about 0.36 to 1.48.}
\label{tab:group_summary}
\resizebox{\linewidth}{!}{%
\begin{tabular}{>{\raggedright\arraybackslash}p{3.6cm}ccccc}
\toprule
\textbf{Group} & \textbf{Mean} & \textbf{SD} & \textbf{Min} & \textbf{Max} & \textbf{Avg Item SD} \\
\midrule
Amount Learned & 2.48 & --   & 2.48 & 2.48 & 1.09 \\
Characters & 3.39 & --   & 3.39 & 3.39 & 1.48 \\
Comparison to Other Apps & 3.10 & 0.46 & 2.51 & 3.74 & 1.12 \\
Fun and Engagement & 3.31 & 0.36 & 2.77 & 3.55 & 1.13 \\
Usability & 3.64 & --   & 3.64 & 3.64 & 1.10 \\
Visuals, Art, and Aesthetic & 3.52 & --   & 3.52 & 3.52 & 1.37 \\
\bottomrule
\end{tabular}%
}
\end{table}

\subsection{Demographic Significance}
We tested two hypothesis for significance:
\begin{enumerate}
\item Native Mandarin users will learn more than users of other native backgrounds since Japanese and Chinese share Kanji characters.
\item Users with a higher level of Japanese will learn more and find the app more fun and engaging compared to beginners of Japanese due to \textit{Jouzu's} immersive approach.
\end{enumerate}

\subsubsection{Native Mandarin User Ratings}
\begin{figure}[h]
\centering
\includegraphics[width=0.9\linewidth]{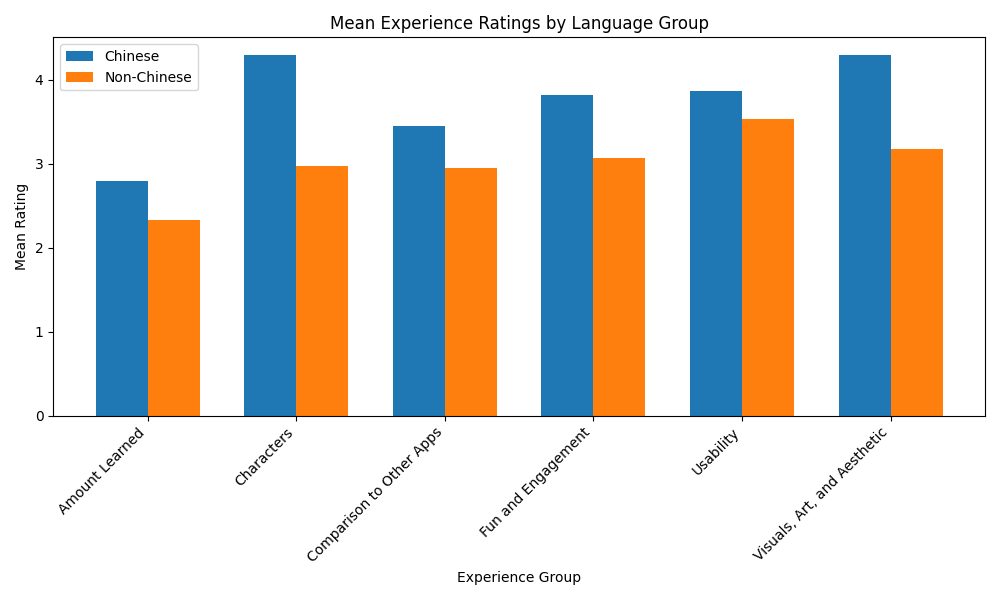}
\Description{A grouped bar chart comparing mean ratings between Chinese native speakers and non-Chinese native speakers. Each category has two bars: the left blue bar for Chinese speakers and the right orange bar for non-Chinese speakers. The figure highlights differences in mean scores across categories, with some categories showing higher ratings among Chinese speakers and others higher among non-Chinese speakers.}
\caption{Mean group ratings for Chinese vs. non-Chinese native speakers. The left bars (blue) represent native Chinese speakers, and the right bars (orange) represent non-Chinese native speakers.}
\label{fig:lang_group_means}
\end{figure}

To examine potential cross-linguistic differences in user experience, participants were grouped by native language: Chinese (including Mandarin) and non-Chinese speakers. Across all six dimensions, Chinese speakers rated the app more favorably. Most notably, the Characters group showed a large gap (\textit{M} = 4.29 vs. 2.97), suggesting strong engagement with stylized agents among Chinese users. Visuals, Art, and Aesthetic also received a much higher rating from Chinese speakers (\textit{M} = 4.29 vs. 3.17), followed by higher scores for Fun and Engagement (\textit{M} = 3.82 vs. 3.07). Usability and Comparison to Other Apps showed smaller but consistent differences, while the largest perceived gap in learning outcomes (\textit{M} = 2.79 vs. 2.33) aligned with open-ended feedback. These results point to meaningful cultural or pedagogical variation in how users interact with character-driven language tools (see Table~\ref{tab:lang_group_means}).
Figure~\ref{fig:lang_group_means} visualizes the mean rating differences between Chinese and non-Chinese participants across all experience groups, highlighting consistently higher ratings by Chinese users.

\begin{table}[h]
\centering
\caption{Mean experience ratings by native language group}
\label{tab:lang_group_means}
\resizebox{\linewidth}{!}{%
\begin{tabular}{>{\raggedright\arraybackslash}p{4.5cm}cc}
\toprule
\textbf{Group} & \textbf{Chinese Mean} & \textbf{Non-Chinese Mean} \\
\midrule
Amount Learned & 2.79 & 2.33 \\
Characters & 4.29 & 2.97 \\
Comparison to Other Apps & 3.45 & 2.95 \\
Fun and Engagement & 3.82 & 3.07 \\
Usability & 3.86 & 3.53 \\
Visuals, Art, and Aesthetic & 4.29 & 3.17 \\
\bottomrule
\end{tabular}%
}
\end{table}

To assess the significance of these cross-linguistic differences, we conducted independent-sample t-tests comparing mean group ratings between Chinese and non-Chinese participants across all six experience categories. As shown in Table~\ref{tab:lang_ttests}, Chinese participants gave significantly higher ratings in the areas of Characters, Fun and Engagement, Visuals, and Comparison to Other Apps. These results suggest that the stylized elements of \textit{Jouzu}, particularly character-driven dialogue and anime aesthetics, may resonate more strongly with learners from Chinese-language backgrounds. While differences in Usability and Amount Learned did not reach statistical significance, both trended in the same direction, indicating a broader pattern of increased satisfaction among Chinese-speaking users.

Based on this significance testing, our first hypothesis was not supported. However, native Chinese speakers performed significantly better for all groups in addition to the amount of learned and usability. That is, native Chinese speakers found the app more fun and engaging, liked the inclusion of anime characters better, preferred the visuals, art, and aesthetic, and found it better overall compared to other language learning apps.

\begin{table}[h]
\centering
\caption{Independent-sample t-tests by native language group. CN = Chinese Language Native, NCN = Non-Chinese Language Native}
\label{tab:lang_ttests}
\resizebox{\linewidth}{!}{%
\begin{tabular}{>{\raggedright\arraybackslash}p{4.5cm}cccccc}
\toprule
\textbf{Group} & \textbf{CN Mean} & \textbf{NCN Mean} & \textbf{t} & \textbf{p} & \textbf{N\textsubscript{CN}} & \textbf{N\textsubscript{NCN}} \\
\midrule
Amount Learned & 2.79 & 2.33 & 1.49 & .1459 & 14 & 30 \\
Characters & 4.29 & 2.97 & 3.27 & .0026 & 14 & 30 \\
Comparison to Other Apps & 3.45 & 2.95 & 2.92 & .0042 & 65 & 149 \\
Fun and Engagement & 3.82 & 3.07 & 4.48 & $<$.0001 & 56 & 120 \\
Usability & 3.86 & 3.53 & 0.98 & .3359 & 14 & 30 \\
Visuals, Art, and Aesthetic & 4.29 & 3.17 & 2.95 & .0059 & 14 & 30 \\
\bottomrule
\end{tabular}%
}
\end{table}

\subsubsection{Differences by Japanese Proficiency Level}

To better understand how language experience shapes user impressions, we analyzed mean group ratings across self-reported Japanese proficiency levels, ranging from complete beginners to near-native speakers. As shown in Table~\ref{tab:jp_proficiency_means}, user satisfaction generally increased with proficiency. Intermediate and native-level users rated Characters and Fun and Engagement highest, both scoring above 3.8 on average. In contrast, complete beginners rated these same categories lower (3.12 and 3.24, respectively), suggesting that more fluent users were better able to engage with stylized dialogue and game-play elements.

A notable difference was observed in the Characters category, where Beginner users reported a sharp increase in enjoyment (\textit{M} = 4.25) compared to Complete Beginners (\textit{M} = 3.12), possibly reflecting increased comprehension of nuanced dialogue. Visuals, Art, and Aesthetic also saw higher scores from more advanced users, echoing trends seen in Figure~\ref{fig:grouped_bar_means}. Overall, these results indicate that \textit{Jouzu} may provide greater value and enjoyment as learners advance in proficiency.

\begin{table}[h]
\centering
\caption{Mean experience ratings by Japanese proficiency level}
\label{tab:jp_proficiency_means}
\resizebox{\linewidth}{!}{%
\begin{tabular}{>{\raggedright\arraybackslash}p{4.3cm}cccc}
\toprule
\textbf{Group} & \textbf{Complete Beginner} & \textbf{Beginner} & \textbf{Intermediate} & \textbf{Native} \\
\midrule
Amount Learned & 2.39 & 2.62 & 3.00 & 3.00 \\
Characters & 3.12 & 4.25 & 4.00 & 4.00 \\
Comparison to Other Apps & 2.96 & 3.42 & 3.80 & 3.80 \\
Fun and Engagement & 3.24 & 3.41 & 3.88 & 3.88 \\
Usability & 3.70 & 3.12 & 4.50 & 4.50 \\
Visuals, Art, and Aesthetic & 3.39 & 3.88 & 4.00 & 4.00 \\
\bottomrule
\end{tabular}%
}
\end{table}

\begin{figure}[h]
  \centering
  \includegraphics[width=0.95\linewidth]{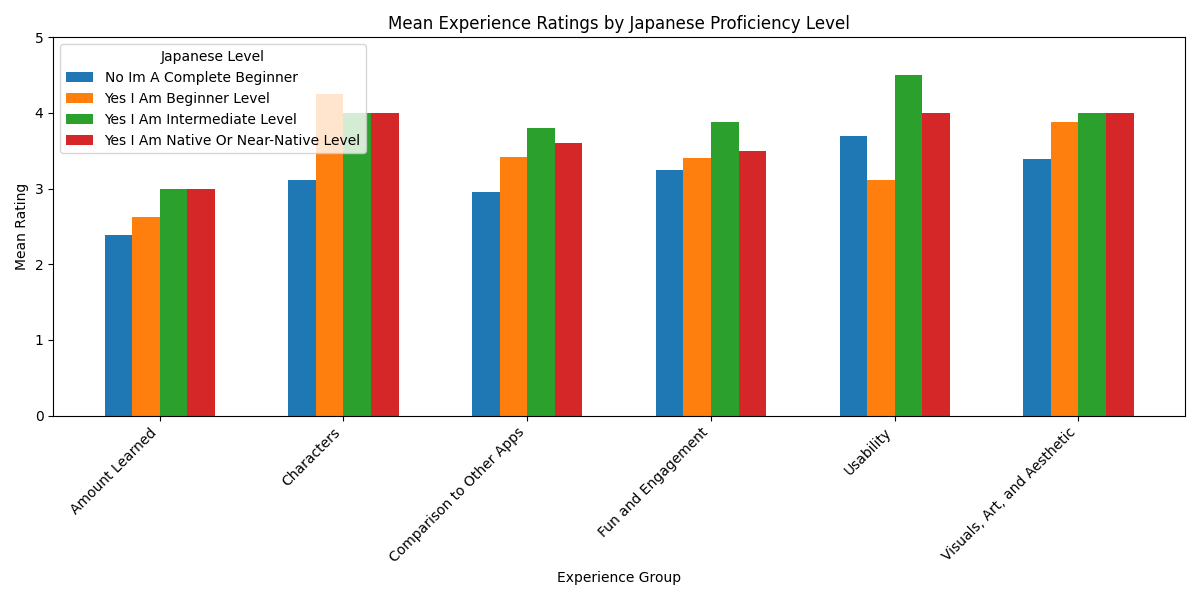}
  \Description{A grouped bar chart showing mean ratings by Japanese proficiency level. Four bars are shown for each category: blue for complete beginners with no exposure, orange for beginners, green for intermediate learners, and red for native or near-native speakers. The chart illustrates how mean ratings vary across proficiency levels, highlighting differences between novices and more advanced or native speakers.}
  \caption{Mean group ratings by Japanese proficiency level. From left to right, the bars represent complete beginner — i.e., the user has had no exposure to Japanese — (blue), beginner level (orange), intermediate level (green), and native or near-native level (red).}
  \label{fig:jp_proficiency_bar}
\end{figure}

\begin{figure}[h]
  \centering
  \includegraphics[width=0.95\linewidth]{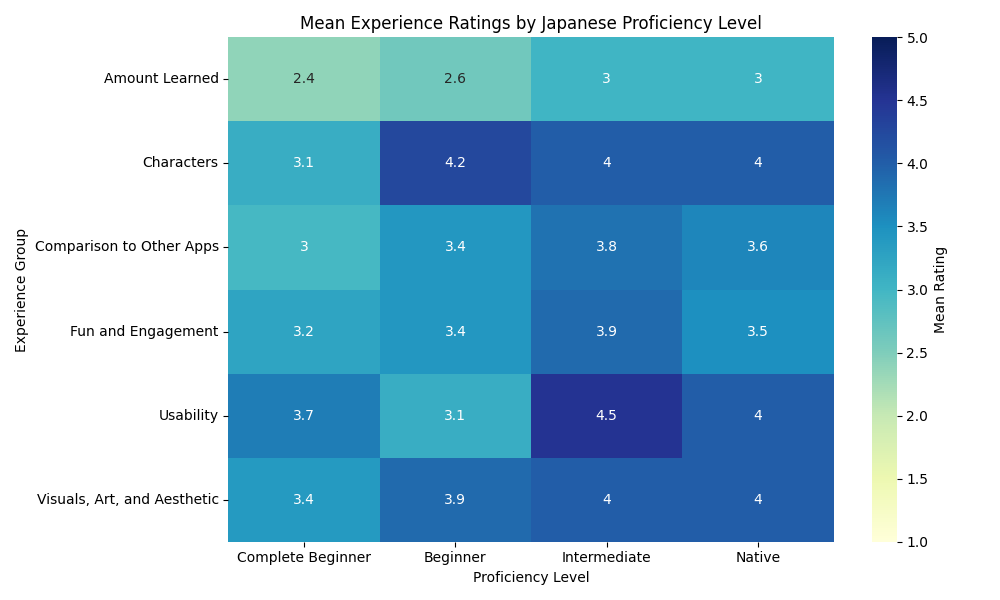}
  \caption{Heatmap of mean experience ratings across Japanese proficiency levels. Darker shades indicate higher user satisfaction per category.}
  \Description{A heatmap showing average user satisfaction ratings for different categories across multiple Japanese proficiency levels.. Darker colors correspond to higher satisfaction ratings, illustrating that more proficient learners reported greater satisfaction in several categories.}
  \label{fig:jp_proficiency_heatmap}
\end{figure}

A heatmap of mean ratings across Japanese proficiency levels (Figure~\ref{fig:jp_proficiency_heatmap}) reveals clear trends in how learners with different backgrounds experienced the app. Ratings for all categories increased with proficiency, but the most pronounced differences appeared in \textit{Characters}, \textit{Fun and Engagement}, and \textit{Visuals, Art, and Aesthetic}. Complete beginners rated the character experience at \textit{M} = 3.12, while intermediate and native-level users both gave it \textit{M} = 4.00 or higher. Similar patterns held for engagement and visual appeal, suggesting that stylized, character-driven interactions resonated more strongly with users who could understand more of the dialogue and cultural nuance. In contrast, usability showed less consistent variation, indicating that learners at all levels found the app interface accessible. These trends underscore the importance of matching language level to interaction design in character-based learning environments.

\section{Qualitative Feedback}
\subsection{Qualitative Feedback by Experience Group}

We conducted a qualitative analysis of open-ended survey responses to four short-answer prompts, using a structured keyword mapping to associate each response with one of six predefined experience categories: \textit{Usability}, \textit{Amount Learned}, \textit{Fun and Engagement}, \textit{Characters}, \textit{Visuals, Art, and Aesthetic}, and \textit{Comparison to Other Apps}. The four prompts asked participants to describe (1) the most helpful feature for learning Japanese, (2) the least helpful feature, (3) the most fun and engaging feature, and (4) any aspects they found frustrating or confusing.

For the question about the most helpful feature, the majority of responses were associated with the \textit{Amount Learned} group. Participants frequently praised the chatbot and quiz mechanics for their ability to reinforce vocabulary and simulate real conversations. One user explained, “The chat feature was honestly the most helpful because it essentially was a chatbot companion for my learning,” while another appreciated being able to ask for phrases and practice pronunciation interactively.

When asked about the \textit{\textbf{least}} helpful feature, responses were again concentrated in the \textit{Amount Learned} group, but with a more critical tone. Users cited difficulty understanding advanced responses in the chat, a lack of scaffolding for beginners, and confusion about how to progress. One participant noted, “You need to have a basis of how to read/write and understand Japanese. Something more simple that starts with the Japanese alphabet would be more helpful.” Usability concerns also surfaced, including confusion in the Learn feature and overwhelming text density in early lessons.

Responses to the most fun and engaging feature were overwhelmingly mapped to the \textit{Fun and Engagement} category. Learners described enjoyment from story interactions, chatting with voiced characters, and the gamified nature of lessons. One user remarked, “It felt like chatting with a friendly sensei (teacher)… It wasn't just study, it felt like real communication.” Others highlighted specific character personalities and the immersive experience of interacting with anime-like avatars.

Finally, in response to frustrations and confusion, feedback was mostly provided for \textit{Usability} and \textit{Characters}. Users reported UI bugs, confusing navigation, lack of direction, and speech recognition issues. One respondent wrote, “I didn’t know how to use the flash cards and couldn’t figure it out,” while another explained that character voices were too quiet and hard to understand. A few users also criticized the app’s difficulty curve, citing challenges understanding kanji and parsing lesson content without sufficient guidance.

Overall, these qualitative results reinforce the quantitative trends observed earlier: participants with more language experience tended to appreciate the app’s depth and character-driven content, while beginners expressed a need for greater structure, onboarding, and transparency in lesson progression.

\subsection{Realized Change in Japanese Proficiency}

\subsubsection{Multiple Choice Performance Analysis}

Participants completed graded Japanese language questions before and after using the app. Prior to usage, only 16 participants attempted the quiz, with no perfect scores recorded. Following app usage, 34 participants completed the quiz -- more than double the original number -- and 2 participants (5.9\%) achieved perfect scores. Additionally, the proportion of high scores (8–10) increased from 12.5\% to 17.6\%. While the proportion of lower scores (1–3) also rose from 31.3\% to 41.2\%, this probably reflects the participation of new learners who gained enough confidence to attempt the quiz but had not yet reached higher proficiency. These results suggest that \textit{Jouzu} supported increased learner engagement and test participation, with gains evident at both the high and entry levels of performance.

\begin{table}[ht]
\centering
\begin{tabular}{|l|c|c|}
\hline
\textbf{Metric (Attempting Users Only)} & \textbf{First Set} & \textbf{Second Set} \\
\hline
Participants Responded & 16 & 34 \\
Perfect Scores (10/10) & 0 (0.0\%) & 2 (5.9\%) \\
High Scores (8--10) & 2 (12.5\%) & 6 (17.6\%) \\
Lower Scores (1--3) & 5 (31.3\%) & 14 (41.2\%) \\
\hline
\end{tabular}
\caption{Comparison of engagement and performance before and after app use (among attempting users)}
\end{table}

\subsubsection{Open Paragraph Assessment}
Participants demonstrated substantial improvements across all three rubric categories: vocabulary diversity, grammatical complexity, and stylistic/pragmatic appropriateness. Group means increased from $M{=}1.75$ to $M{=}3.25$ for both vocabulary and grammar, and from $M{=}2.25$ to $M{=}4.00$ for style. Average gains were $+1.50$ (vocabulary), $+1.50$ (grammar), and $+1.75$ (style). Standard deviations narrowed post-intervention (e.g., vocabulary $SD{=}1.98 \rightarrow 1.28$), suggesting more consistent performance. Effect sizes were large in all cases (paired Cohen's $d = 0.85$--$1.06$).

\paragraph{User-level improvements.}
Five of eight users showed measurable improvements in all three categories (Users 1, 2, 4, 5, and 7). User~5 exhibited the largest overall gain, improving from a total of 0 to 13 points ($\Delta{=}+13$). User~7 improved from 0 to 9 ($\Delta{=}+9$); User~1 from 0 to 7 ($\Delta{=}+7$); User~2 from 9 to 15 ($\Delta{=}+6$); and User~4 from 0 to 3 ($\Delta{=}+3$). Users~3, 6, and 8 showed no changes (totals of 15, 11, and 11 respectively, before and after) See Table \ref{tab:rubric-results} for rubric scores.

\paragraph{Distribution and relationships.}
At the category level, five participants improved in vocabulary, five in grammar, and five in style; no decreases were observed. Improvements were highly correlated (e.g., vocabulary--style $r{=}0.94$; vocabulary--grammar $r{=}0.80$; grammar--style $r{=}0.92$). Lower starting scores predicted larger gains (e.g., vocabulary start vs.~improvement $r{=}-0.76$), suggesting that the intervention disproportionately benefited participants who began with weaker performance. Overall rubric sums increased from 46 to 84 ($\Delta{=}+38$). See Table \ref{tab:user-totals} for per-user totals.

% ---------------------------
% Table 1: Group-level stats (resizebox version)
% ---------------------------
\begin{table}[h]
\centering
\caption{Rubric scores before and after the intervention ($N{=}8$). The labels are abbreviated as BM (before-intervention mean), AM (after-intervention mean), standard deviations (SD), IM (improvement mean), which is defined as (After$-$Before), and paired Cohen’s $d$.}
\label{tab:rubric-results}
\resizebox{\linewidth}{!}{%
\begin{tabular}{lcccc}
\toprule
\textbf{Category} & \textbf{BM (SD)} & \textbf{AM (SD)} & \textbf{IM} & \textbf{Cohen's $d$} \\
\midrule
Vocabulary Diversity & 1.75 (1.98) & 3.25 (1.28) & +1.50 & 1.06 \\
Grammatical Complexity & 1.75 (1.98) & 3.25 (1.67) & +1.50 & 0.85 \\
Stylistic / Pragmatic App. & 2.25 (2.49) & 4.00 (1.51) & +1.75 & 0.96 \\
\midrule
\textbf{Overall (sum)} & 46 & 84 & +38 & -- \\
\bottomrule
\end{tabular}%
}
\end{table}

% ---------------------------
% Table 2: User-level totals
% ---------------------------
\begin{table}[ht]
\centering
\caption{Per-user totals (sum of the three categories) before and after the intervention, with deltas.}
\label{tab:user-totals}
\begin{tabular}{lccc}
\toprule
\textbf{User} & \textbf{Before Total} & \textbf{After Total} & \textbf{Delta} \\
\midrule
User 1 & 0  & 7  & +7 \\
User 2 & 9  & 15 & +6 \\
User 3 & 15 & 15 & 0 \\
User 4 & 0  & 3  & +3 \\
User 5 & 0  & 13 & +13 \\
User 6 & 11 & 11 & 0 \\
User 7 & 0  & 9  & +9 \\
User 8 & 11 & 11 & 0 \\
\midrule
\textbf{Totals} & \textbf{46} & \textbf{84} & \textbf{+38} \\
\bottomrule
\end{tabular}
\end{table}

\paragraph{Paragraph Examination} In addition to quantitative scores, we examined the paragraphs produced by participants to better understand the nature of their progress. These qualitative samples illustrate differences in fluency, grammatical control, and stylistic choices across proficiency levels. For instance, an intermediate speaker wrote the following:
\begin{CJK}{UTF8}{min}
\begin{quote}
私は日本語を勉強しています. 最初はひらがなとカタカナを覚えるのが難しかったです.でも, 毎日少しずつ練習して, 今は簡単な会話ができます. 将来,日本に旅行して, 実際に日本語を使ってみたいです.
\end{quote}
\begin{quote} Translation: I am studying Japanese. At first, it was difficult to memorize Hiragana and Katakana. But by practicing little by little every day, I can now have simple conversations. In the future, I want to travel to Japan and try using Japanese in real life.
\end{quote}
\end{CJK}

The paragraphs from participants who were unable to write them before had varying degrees of fluency. For example, one student who identified as a beginner wrote the following:
\begin{CJK}{UTF8}{min}
\begin{quote}
    私の日本語はまだ初歩的ですが、食べ物などについて書けるようになりました。うどんは歯ごたえがあってスープが美味しいので大好きです。近いうちに日本に行くのが楽しみです。大学卒業後かな。
\end{quote}
\begin{quote}
    Translation: My Japanese is still basic, but I've become able to write about things like food. I love udon because it has a nice chewy texture and the soup is delicious. I'm looking forward to going to Japan sometime soon—maybe after I graduate from university.
\end{quote}
\end{CJK}

Others demonstrated basic vocabulary acquisition, not limited to the vocabulary words used in the character's individual language styles.
\begin{CJK}{UTF8}{min}
\begin{quote}
    konichiwa, ogenki desu ka. Sumimasen, doko toire desu ka. Sore nan desu ka. Arigatou. Naze (why), itsu (when), dare (who).
\end{quote}
\begin{quote}
    Translation: Hello, how are you? Excuse me, where is the bathroom? What is that? Thank you.
    Why, when, who.
\end{quote}

Another student: 
\begin{quote}
    おはようございます- お元気ですか (My attempt at "good morning, how are you?")
\end{quote}
\end{CJK}

One participant who was already identified as a beginner with some experience did not attempt a paragraph before use but constructed a comprehensive sentence after use. It is important to note that, while characters do not use polite Japanese, this student still wrote a sentence using polite Japanese mixed with Saya's character language.

\begin{CJK}{UTF8}{min}
\begin{quote}
    私は武道が好きです
\end{quote}
\begin{quote}
    Translation: I like bushi
\end{quote}
\end{CJK}

\subsection{Chatbot Interaction}

We observed diverse interaction strategies among users engaging with \textit{Jouzu’s} character-based chat feature. Across early-stage usage, three recurring patterns emerged, particularly among beginners.

First, several participants initiated conversations in English by asking the characters to introduce a Japanese word. In response, characters typically presented a common word or phrase, accompanied by a simple explanation in Japanese. For example, one user typed, “Teach me a word in Japanese,” prompting a character to reply with an entry-level greeting and its meaning. These interactions provided accessible entry points into vocabulary acquisition, even for those with no prior exposure to Japanese.

Second, users frequently asked how to translate specific English words into Japanese, such as “How do you say sword in Japanese?” Character responses often included the target term in Kanji with furigana or spoken pronunciation, demonstrating how the system scaffolds lexical learning through bilingual dialogue. This behavior echoes documented beginner strategies in computer-assisted language learning, where learners rely on translation-based queries to build foundational vocabulary.

Third, a subset of participants conducted simple, English-initiated conversations about everyday topics. The characters responded entirely in Japanese, allowing users to passively absorb sentence structure and phrasing. For instance, a user might ask, “What is your favorite tea?” and receive a culturally contextualized response in natural-sounding Japanese. These exchanges continued smoothly, with characters maintaining conversational coherence even when users offered minimal or English-only input.

In contrast, intermediate and advanced users tended to rely more heavily on Japanese in their chat entries. Many incorporated full sentences and questions in Japanese, using the characters to simulate real-time dialogue in a stylized but grammatically accurate register. Notably, more fluent users also engaged with the speech recognition functionality, using spoken input to practice pronunciation and listening comprehension. These behaviors suggest that the chat system offers scalable interaction, adapting fluidly to user proficiency levels. While beginners benefit from bilingual support and forgiving prompts, advanced users appear to leverage the system as a semi-authentic environment for spoken Japanese practice.

\section{Results}

We conducted a mixed-methods evaluation of user interaction with \textit{Jouzu}, a multimodal language learning environment featuring stylized, voiced conversational character agents. Participants rated the system most highly for usability (\textit{M} = 3.64), visual and aesthetic appeal (\textit{M} = 3.52), and character engagement (\textit{M} = 3.39), indicating strong positive appreciation of both the interface design and affectively rich agent behavior. Although perceived learning outcomes received lower scores on average (\textit{M} = 2.48), open-ended responses suggest that users viewed \textit{Jouzu} primarily as a motivational, immersive supplement to traditional instruction, rather than a standalone pedagogical solution.

Demographic analysis revealed consistent trends. Native Chinese speakers rated the system significantly higher across multiple experience dimensions, particularly for visual appeal, character interaction, and overall engagement. This pattern suggests stronger alignment with the anime-inspired aesthetic and the affective realism of stylized agent dialogue. Additionally, participants with higher self-reported Japanese proficiency gave more favorable ratings across all categories. Intermediate and native-level users in particular reported greater enjoyment of character language, voiced responses, and narrative immersion, reinforcing the utility of stylization at more advanced stages of language learning.

Analysis of interaction logs further revealed different usage strategies by proficiency level. Complete beginners commonly used English prompts to request translations or vocabulary (e.g., “What is ‘sword’ in Japanese?”), with agents responding in Japanese using character-specific phrasing. These users often relied on passive listening or hybrid input patterns. In contrast, intermediate and advanced users engaged in full-sentence Japanese conversations, using both text and speech input to simulate native dialogue interaction.

Participants completed graded Japanese language questions before and after using the app. Prior to usage, only 16 participants attempted the quiz, with no perfect scores recorded. After app usage, 34 participants -- more than double the original number -- completed the quiz, with 2 participants (5.9\%) achieving perfect scores. The proportion of high scores (8–10) also increased from 12.5\% to 17.6\%, indicating vocabulary gains for some learners. While the percentage of lower scores (1–3) rose from 31.3\% to 41.2\%, this likely reflects new users gaining enough confidence to attempt the quiz despite not yet achieving mastery. These results suggest that \textit{Jouzu} fostered both increased test participation and broader performance distribution, with gains at the upper end and expanded access at the entry level.

Writing samples submitted before and after app usage revealed meaningful expressive gains, particularly among novice learners. Several users who were unable to produce any Japanese output prior to using the app submitted short paragraphs afterwards, indicating increased confidence and productive vocabulary. These paragraphs included both basic grammatical constructions and character language vocabulary, suggesting exposure to stylistic features embedded in the agent dialogue.

In addition to these broad trends, our rubric analysis shows consistent patterns of improvement across participants. On average, scores increased by 1.5 points in vocabulary and grammar, and by 1.75 points in stylistic and pragmatic appropriateness, all with large effect sizes. Five of the eight participants improved in every category, with gains especially pronounced among those who began with lower initial scores. By contrast, participants who entered the study with higher baseline proficiency tended to maintain their levels rather than progress further. This suggests that the system was particularly effective in supporting learners who started with limited skills, while more advanced users primarily experienced consolidation rather than measurable growth.

Taken together, these findings demonstrate that \textit{Jouzu’s} stylized conversational agents foster high levels of engagement and perceived usability across a diverse user base. While advanced learners derived the most direct value from immersive dialogue, even complete beginners demonstrated affective benefits and signs of early language uptake.

\section{Discussion}

Our findings suggest that stylized, voiced dialogue agents can significantly enhance user engagement and motivation in computer-assisted language learning. While \textit{Jouzu} offers multiple learning features including lessons, flashcards, and a sentence-level inspector, the emotionally expressive agents were consistently cited as the most engaging component. Users reported that the character voices, visual design, and in-character phrasing created a more immersive and socially meaningful learning experience than traditional tools.

However, stylization alone was not sufficient for all users. Beginners often struggled to understand the agents’ responses without more structured support, highlighting the need for adaptive scaffolding and clearer progression cues. In contrast, intermediate and advanced learners reported higher enjoyment and made fuller use of the conversational agents, sometimes mirroring their stylized language. These patterns reflect existing work on personalization in dialogue and CALL systems and show that character-based interaction may be most effective when aligned with the learner’s proficiency.

Multiple choice test results indicate both an increase in participation and signs of improvement at the top end of performance. After using the app, more users completed the quiz, and a greater proportion achieved high scores (8–10) or perfect results. While some users also scored in the lower range, this likely reflects the broader base of participants, including those who previously may not have attempted the test. These findings suggest that \textit{Jouzu} encouraged learners across a range of proficiency levels to engage with structured assessment, and that some users experienced measurable vocabulary gains. Importantly, these outcomes complement the expressive growth observed in paragraph writing, reinforcing the system’s effectiveness in supporting both test-oriented recall and open-ended language use.

\textit{Jouzu} emphasizes open-ended, affective interaction over rote memorization, and several users who showed weaker multiple choice performance nonetheless demonstrated stronger expressive ability in paragraph writing. Many participants who initially wrote little or no Japanese were later able to produce multi-sentence output using vocabulary, grammar, or stylistic features encountered in the app. This suggests that the app supports exploratory, confidence-building learning, which may not be fully captured by short-format assessments.

Overall, \textit{Jouzu’s} design suggests that culturally stylized, emotionally expressive dialogue agents -- when embedded in a multimodal learning environment -- can drive engagement and support meaningful language practice. Future work should explore adaptive features to help beginner users benefit more fully from the system’s immersive elements.

\subsection{Design Implications}

The findings from this study suggest several concrete design directions for future conversational language learning agents. The incorporation of stylized agents can serve not just as a novelty, but as a mechanism for increasing user motivation and affective immersion, particularly among learners with intermediate or higher proficiency. These users often mirrored stylistic elements in their own output and remained engaged longer during conversational sessions. However, the same features that support expressive language play for advanced users can present barriers for novices. Accordingly, systems may benefit from adaptive scaffolding mechanisms that calibrate agent responses based on learner proficiency. For instance, stylization could be made toggleable or modified dynamically during the session to scaffold more accessible interaction while retaining affective richness.

\begin{itemize}
\item Stylized dialogue agents can increase motivation and expressive risk-taking, particularly for intermediate learners.
\item Beginners need scaffolding to navigate character language; toggleable stylization or adaptive agent personas may improve accessibility.
\item Cultural alignment (e.g., anime style for East Asian learners) boosts affective engagement, but may limit generalizability.
\end{itemize}

\subsection{Limitations}

This study has several limitations that should be taken into account when interpreting results or attempting to generalize findings. First, the participant pool consisted of computer science students, many of whom already had familiarity with technology or anime aesthetics, which may not reflect broader learner populations. Second, participation in assessments was optional, and baseline proficiency data was self-reported, limiting control over baseline variation and motivation. Third, while stylization proved effective for our sample, it was tightly tied to Japanese media tropes; different cultural stylizations may be needed to engage other learner communities. The absence of a control group also limits causal claims about effectiveness relative to traditional methods. Finally, our evaluation of open-ended writing relied on a manually applied rubric rather than a standardized or automated metric, though we explicitly justified this approach for its categorical clarity.

\section{Conclusion}
\textit{Jouzu} demonstrates that stylized, character-based interaction enhances engagement, confidence, and cultural immersion, and that this engagement aligns with measurable learning improvements. After use, participants showed increases at the upper end of multiple-choice performance and substantial gains in expressive writing (vocabulary diversity, grammatical complexity, and stylistic appropriateness), alongside broader participation in assessment. These results suggest that culturally resonant, affective design can transform engagement into practice that supports learning, while complementing, rather than replacing, structured instruction. Future work should add adaptive scaffolding for beginners and examine how sustained engagement with stylized agents impacts longer-term spoken fluency.

% \section*{Acknowledgements}
% This research was initiated independently of the primary author’s academic responsibilities at Columbia University and was conducted without the use of university resources.

\begin{CJK}{UTF8}{min}
\bibliographystyle{ACM-Reference-Format}
\bibliography{sample-base}
\appendix
\section{Survey Questions}
\label{appendix:survey_questions}
\subsection{Pre-use Questions}
The users were introduced to the survey with the following introduction.
\begin{quote}
\ttfamily
    Welcome to my study! This study is a user study for Jouzu, a mobile application for learning Japanese with original anime-inspired characters. This survey will ask for basic information regarding your device, and a 10-15 minute assessment in Japanese. Your UNI (for CU students) and/or email (if external) will be tracked for assignment credit purposes. All data will be anonymized for analysis and publication. Please note that refusal to participate will have no influence on your academic standing.

By signing up for this user test and creating an account on the app, you agree to our Terms of Service.
\end{quote}
The users were asked the following background information questions.

\begin{quote}
\ttfamily
\begin{enumerate}
    \item Have you studied Japanese before? (It's completely okay if you haven't)
    \item What is your native language?
    \item Which other languages do you speak? (You don't have to be completely fluent in them, but at least have a basic understanding).
    \item What is your age?
    \item Which (if any) of these apps have you used before?
    \item Which (if any) of these games have you played?
    \item Which (if any) of the following games have you played?
\end{enumerate}
\end{quote}

\section{Appendix: App Evaluation Questions}
\label{appendix:evaluation_questions}
Users were introduced to the post-survey with the following introduction.
\begin{quote}
\ttfamily
    Thank you very much for using my app! Here, you will be asked several questions about your experience, then take another short Japanese assessment. Please answer all questions to the best of your ability this time, but don't panic if you don't know any answers. There is no expectation for your score.
\end{quote}
The following questions were asked to evaluate the application.
\begin{quote}
\ttfamily
\begin{enumerate}
    \item How long did you use the app?
    \item Which features did you use?
    \item How engaging was the app overall?
    \item How fun was it to use?
    \item How much did the app help you learn or practice Japanese?
    \item How motivated did you feel to keep using it?
    \item How easy was the app to use?
    \item Which feature was the most helpful for you to learn and practice Japanese?
    \item Which feature was the least helpful for you to learn and practice Japanese?
    \item Which feature was the most frustrating or confusing?
    \item Which feature was the most fun and engaging?
    \item How much did the music improve your experience?
    \item How much did the animations enhance your engagement?
    \item How much did the anime-style visuals make the app more appealing?
    \item How much did the characters enhance your learning experience?
    \item Which of these aesthetic elements stood out the most? (Short answer)
    \item Which character did you enjoy chatting with the most?
    \item Why did you prefer that character? (Short answer)
    \item Did the characters make the app more enjoyable overall? \\
    1 = not at all, 2 = slightly, 3 = somewhat, 4 = moderately, 5 = extremely
    \item Compared to other language learning apps, how would you rate the app in terms of fun? \\
    1 = much worse, 2 = slightly worse, 3 = the same, 4 = slightly better, 5 = much better
    \item Compared to other language learning apps, how would you rate the app in terms of helpfulness for learning? \\
    1 = much worse, 2 = slightly worse, 3 = the same, 4 = slightly better, 5 = much better
    \item Compared to other language learning apps, how would you rate the app in terms of ease of use? \\
    1 = much worse, 2 = slightly worse, 3 = the same, 4 = slightly better, 5 = much better
    \item Compared to other language learning apps, how would you rate the app in terms of your motivation to keep learning? \\
    1 = much worse, 2 = slightly worse, 3 = the same, 4 = slightly better, 5 = much better
    \item Compared to other language learning apps, how would you rate the app in terms of visual and auditory style? \\
    1 = much worse, 2 = slightly worse, 3 = the same, 4 = slightly better, 5 = much better
    \item What did you like most about the app?
    \item What did you like least or find frustrating/ \\ confusing?
    \item What would you change about the app?
    \item Any suggestions for improvement?
\end{enumerate}
\end{quote}

\section{Pre- and post-use Japanese Quiz Questions}
\label{appendix:quiz_questions}

\subsection{Pre-use Quiz}
The users were introduced to the Japanese quiz with the following explanation.
\begin{quote}
\ttfamily
Here, you will be asked a few questions to assess your Japanese knowledge. The results of this test will be compared against another test after you finish using the app.

If you have not studied Japanese before or are not confident in taking a Japanese test, please skip this part and submit the form. Skipping this part will not affect your participation or assignment credit in any way.
 
For those who decide to continue, if you don't know an answer, feel free to skip the question or take your best guess. Some of these questions will be very difficult and may be beyond your current knowledge, so do not stress about getting a high score.
\end{quote}
The users were asked the following questions.
\begin{quote}
\ttfamily
    \begin{enumerate}
    \item 「しんしん」を漢字で書いてください。
    \item 紡ぎ（つむぎ）の意味を選びなさい。
    \item 「じひ」を漢字で書いてください。
    \item 友達（ともだち）とは何ですか？
    \item 語らい（かたらい）とは何ですか？
    \item 新しいくるまですね
    \item （　　）しないで、どうぞたくさんたべてください。
    \item 3日前（みっかまえ）から雨（あめ）が続いている。
    \item 事故の原因は、機械の（　　）作動にあると考えられている。
    \item 自分が未熟すぎて心配だなどというのは、未熟ということをマイナスに考える証拠だ。（　　）弱い人間とか未熟な人間のほうが、はるかにふくれあがる可能性をもっている。
    \item Please write 1 paragraph (at least 4 sentences) in Japanese about any topic you would like.
\end{enumerate}
\end{quote}

\subsection{Post-use Quiz}
The users were introduced to the post-use quiz with the following instructions.
\begin{quote}
\ttfamily
    Please answer as many questions as you can to the best of your ability. None are required, however, so do not feel pressured to answer them if you do not know the answers.
\end{quote}

The following quiz questions were asked to the user.
\begin{quote}
\ttfamily
\begin{enumerate}
    \item 精神（せいしん）とは何ですか？
    \item 名誉を読むとき、正しいふりがなは何ですか？
    \item 崇高（すうこう）に最も近い英語の意味を選んでください。
    \item 自然観を読むとき、正しいふりがなは何ですか？
    \item 歴史（れきし）の意味を選んでください。
    \item きのうはうちに（　　）何をしましたか。
    \item 今日は、ゴミのしゅうしゅう日ですか。
    \item わたしのせんもんは医学です。
    \item 登山（とざん）の途中（とちゅう）で、（　　）と思いました（おもいました）。なぜかというと、予想（よそう）よりも山（やま）の道（みち）を歩く（あるく）のは大変（たいへん）だったからです。
    \item おもちゃというと、ただ子供が遊ぶためだけのものだと（　　）。
    \item Please write 1 paragraph in Japanese about any topic you would like.
\end{enumerate}
\end{quote}

\section{Open-ended Paragraph Grading Rubric}
\label{appendix:rubric}
The following rubric was applied for grading open-ended paragraph responses.

{\ttfamily
Analytic rubric for Japanese writing samples (1--5 scale)

\textbf{Vocabulary Diversity}
\begin{enumerate}
  \item[1:] Very small pool of high-frequency words; heavy repetition.
  \item[3:] Mix of common items with some topic-specific terms; moderate repetition and some paraphrase.
  \item[5:] Broad range of lexical items including topic-specific and stylistic lexemes; paraphrase and word-family variation evident.
\end{enumerate}

\textbf{Grammatical Complexity}
\begin{enumerate}
  \item[1:] Isolated words/fragments or only simple SOV sentences; minimal particle variety.
  \item[3:] Simple sentences plus occasional coordination/subordination (e.g., ～て, ～から); some clause chaining; emerging particle control.
  \item[5:] Varied clause combining (relative/embedded clauses, conditionals, modality/aspect); consistent and context-appropriate particle usage.
\end{enumerate}

\textbf{Stylistic / Pragmatic Appropriateness}
\begin{enumerate}
  \item[1:] No discernible register/persona; inconsistent or inappropriate politeness; character-language markers absent.
  \item[3:] Some stylistic markers (e.g., sentence-final particles, limited persona cues); partly consistent with task and audience.
  \item[5:] Sustained register management and persona (e.g., pronouns, sentence-final forms, particles) appropriate to task/character language; stylistic choices enhance communicative intent.
\end{enumerate}
}

\end{CJK}

% Notes for readers: dimensions align with CAF (complexity/accuracy/fluency) and CEFR sociolinguistic appropriateness;
% lexical diversity and syntactic complexity have validated operationalizations in L2 writing research.

\end{document}